\documentclass[showpacs,prb,letterpaper,twocolumn,aps,epsf]{revtex4}
\usepackage{dcolumn}
\usepackage{amsmath,amssymb}
\usepackage{graphicx}

\newcommand{\im}{\mathrm{i}}
\newcommand{\ex}[1]{\mathrm{e}^{#1}}
\newcommand{\du}{\mathrm{d}}

\newcommand{\ve}[1]{\mathbf{#1}}

\begin{document}
\title{
Effects of boundaries and density inhomogeneity on states of vortex 
matter in Bose--Einstein condensates at finite temperature
}

\author{S. Kragset${}^1$, E. Babaev${}^{2,3}$, and A. Sudb\o ${}^{1,4}$}

\affiliation{ ${}^1$ Department of Physics, Norwegian University
  of Science and Technology, N-7491 Trondheim, Norway \\
  ${}^2$ Physics Department, University of Massachusetts, Amherst,
  01003-9337, USA \\
  ${}^3$ Department of Theoretical Physics, The Royal Institute of
  Technology 10691 Stockholm, Sweden \\
  ${}^4$ Center for Advanced Study, The Norwegian Academy of Science and Letters,
  Drammensveien 78, 0271 Oslo, Norway}

\begin{abstract}
  Most of the literature on quantum vortices predicting various states
  of vortex matter in three dimensions at finite temperatures in
  quantum fluids is based on an assumption of an extended and
  homogeneous system.  It is well known not to be the case in actual
  Bose--Einstein condensates in traps which are finite systems with
  nonuniform density. This raises the question to what extent one can
  speak of different aggregate states of vortex matter (vortex
  lattices, liquids and tensionless vortex tangle) in these system. To
  address this point, in the present work we focus on the finite-size,
  boundaries and density inhomogeneity effects on thermal vortex
  matter in a Bose--Einstein condensate.  To this end we perform Monte
  Carlo simulations on a model system describing trapped
  Bose--Einstein condensates.  Throughout the paper, we draw on
  analogies with results for vortex matter obtained for extended
  systems. We also consider, for comparison, the cylindrical
  confinement geometry with uniform density profile from the center
  out to the edge of the trap. The trapping potential is taken to be
  generically of an anharmonic form in such a way as to interpolate
  between a harmonic trap and a cylindrical confinement geometry. It
  also allows for a dip in the density profile at the center. We find
  distinct thermal equilibrium properties of the vortex system as the
  qualitative characteristics of the trapping potential is varied. For
  a uniform cylindrical confinement geometry, we find a vortex lattice
  at the center of the trap as well as ring-like thermal fluctuations
  of the vortex system as the trap edge is approached. For a harmonic
  trap, we find a distinct region at the edge of the trap where the
  vortex lines appear to have lost their line tension. Due to the
  steep density gradient, this crosses directly over to a vortex-line
  lattice at the center of the trap at low temperatures. At higher
  temperatures, an intermediate tensionful vortex liquid may
  exist. For an anharmonic trap where the ground state condensate
  density features a local minimum at the center of the trap, we find
  a corresponding region which appears to feature a tensionless
  vortex-line liquid phase. This work suggests that 
  finiteness and intrinsic inhomogeneity of the system not withstanding,
  one nonetheless can approximately invoke the notion of distinct aggregate 
  states of vortex matter realized at certain length scales. This might be 
  helpful, in particular in search of possible new states of vortex matter 
  in Bose--Einstein condensates with multiple components and different 
  symmetries.
\end{abstract}


\maketitle 

\section{\label{sec:intro} INTRODUCTION}
The physics of trapped gases on the one hand, and the physics of
superconductors and superfluids on the other, may be conceptually
linked by rotating Bose--Einstein condensates (BEC) in magnetic traps
\cite{Coddington_Thesis}, or pair condensates of ultracold Fermi gases
\cite{Greiner_Thesis,Regal_Thesis,Zwierlein_Thesis}.  Superconductivity,
superfluidity, and Bose--Einstein condensation are all hallmarks of
quantum fluids governed by the onset of long-range phase coherence in
a macroscopic matter wave.  This phase-coherence is the matter-wave
analog of a corresponding well-known phenomenona in electromagnetic
waves, namely phase-coherence in such waves established by stimulated
emission of radiation. One distinguishing feature of such quantum
fluids is that the macroscopic matter wave function is complex
$\psi(r) = |\psi(r)| e^{i \theta(r)}$, and with a phase $\theta(r) \in
[0, 2 \pi \rangle$, i.e. this phase is defined with a compact
support. This has far-reaching ramifications for the physics in the
sense that the order parameter of the system supports stable
topological defects in the form of point-vortices in two dimensions
and vortex-loops and vortex-lines in three dimensions. Phase
transitions from superfluids to normal fluids, are entirely governed
by such topological defects.
\par
In two dimensions, this is manifested in the well known
Kosterlitz--Thouless phase transition of unbinding of a
vortex-antivortex pair from a low-temperature superfluid with only
tightly bound vortex-antivortex pairs to a normal fluid in a process
where the most weakly bound pairs are dissociated \cite{KT}. It is
also equivalent to a phase transition from a dielectric to a metal in
the two-dimensional Coulomb-gas with overall charge-neutrality
\cite{Hoye_Olaussen}. In three dimensions, one has a phase-transition
from an ordered low temperature phase with at most a few small closed
vortex loops present, to a normal fluid where closed vortex loops have
proliferated throughout the system into a tangle in such a way as to
make it possible to connect opposite sides of a macroscopic system
with a connected vortex path
\cite{Kleinert1982,Tesanovic1999,NguyenSudbo1998,NguyenSudbo1999,Kleinert_book,Fossheim_Sudbo_book}.
This was originally proposed by Onsager already in 1949 in qualitative terms as a way of explaining 
the $\lambda$-transition in superfluid $^4$He \cite{Onsager_vortexloop}.  In the context of the present 
paper, we stress that while Refs. \onlinecite{Onsager_vortexloop,Kleinert1982} dealt with the issue for 
zero rotation for the superfluid or zero magnetic field for the superconductor, the picture was extended 
to finite rotation in superfluids and finite magnetic field in extreme type-II superconductors in
Refs. \onlinecite{Tesanovic1999,NguyenSudbo1998,NguyenSudbo1999}. It is the latter situation that is 
relevant to rotating Bose--Einstein condensates.
\par
The connectivity of the thermally excited vortex tangle changes
abruptly at the critical temperature of the system. Such a phase
transition in three dimensions cannot be cast into the framework of a
Kosterlitz--Thouless phase transition for the simple reason that
vortex loops in three dimensions have a property that
vortex-antivortex pairs in two dimension do not have, namely
shape. This contributes significantly to the configurational entropy
of the system. The fact that vortex loops can have extremely
complicated geometric shapes and will form a fractal structure at long
length scales at the critical point, is crucial in order for them to
be able to proliferate. It also means that the vortex loops cannot be
regarded as renormalized vortex rings with a 'doughnot hole' in the
middle. They are instead fractal objects with fractal dimension
$D_H \approx 2$ \cite{HoveSudbo}. Such a fractal dimension is considerably 
larger than what it would have been for ring-like objects, namely $D_H=1$. 
The entropy production associated with this proliferation of topological 
defects is accompanied by a loss of a generalized stiffness, in this case 
the superfluid density or phase stiffness of the system. In this sense,
the above scenarios both in $2$D and $3$D fall nicely within a general
definition of a phase transition, namely that a phase transition
occurs at a point where some generalized stiffness is lost as a result
of a spontaneous proliferation of stable topological defects of the
complex scalar matter fields in the system \cite{PWABasicNotions}.
\par
Over the last decade, remarkable progress has been made in achieving
Bose--Einstein condensates in gases of ultra-cold atoms in a magnetic
trap \cite{Ketterle_Nobelwork,Cornell_Nobelwork,Rice_work}. Such
condensates are now being routinely manipulated in a large variety of
ways, and may for instance be spun up to produce vortex lattices of a
condensate in a magnetic trap \cite{BECVL1,BECVL2}. One may also
envisage low-dimensional vortex structures \cite{1dPRL,1dPRA}.  There
are even cases where other aggregate states of vortex matter are known to
exist in quite different condensed matter systems, such as two-component 
superconductors with individually conserved condensates \cite{NWA2004}. It 
is the purpose of this paper to study the thermal excitations of
vortex-lines and vortex-loops in Bose--Einstein condensates which are
confined to a cylindrical geometry by a trapping potential. This
trapping potential generally increases from the center of the trap
towards the edge of the condensate, although this variation may be 
non-monotonic. In essence, it acts as a spatially
dependent chemical potential for the bosons in the system, thus
affecting the overall condensate density. The density is typically highest 
at the center of the trap and vanishes towards the edge of the trap,
although more complicated profiles may easily be envisaged, and will
in fact be considered in this paper. The
way in which the overall density varies is directly determined by the
trapping potential. Thus, such systems are inherently nonuniform and
therefore it is important specifically to study the effect of spatial
density variations.
\par
This paper is organized as follows. In Section \ref{sec:model}, we
motivate and introduce the model on which we will perform Monte Carlo
simulations in this paper. In Section \ref{sec:montecarlo}, we
describe the Monte Carlo simulations we will perform. In Section
\ref{sec:uniform} we present results for a uniform system, as a
benchmark, and we also consider such a system confined in a
cylindrical geometry. The former of these results are applicable to
fluctuating vortex matter in extreme type-II superconductors, where we
can neglect the fluctuations in the gauge field. The latter results
apply to rotating helium in a container. In Section
\ref{sec:nonuniform} we present our results for two different types of
traps, namely the harmonic cylindrical trap, and the anharmonic
cylindrical trap. The quantities we focus on are the helicity
modulus and structure functions of the vortex matter in the systems in
the various parts of the confinement and traps, i.e. at various
distances from the center. In Section \ref{sec:conclusion} we present 
our conclusions. This work is a follow-up of the letter 
Ref. \onlinecite{Kragset}.

\section{\label{sec:model} GENERAL MODEL}
We will use Monte Carlo (MC) computations to search for the thermal
equilibrium vortex states of a rotating Bose--Einstein condensate in
three dimensions. A natural starting point is the Gross-Pitaevskii
free energy functional in a rotating frame, given by the energy
functional
\begin{equation}
  \label{eq:gpenergy}
  E' = \int dV \Bigl[\psi^{\ast}(-\frac{\hbar^2}{2 M} \nabla^2 + V(r) -
  \Omega \hat{L}_z)\psi + \frac{1}{2} g |\psi|^4\Bigr].
\end{equation}
Here, $\hat{L}_z = \im \hbar (y \partial_x - x \partial_y)$ and $g = 4
\pi a_s \hbar^2 / M$, $M$ is the mass of the  atoms in the condensate,
and $a_s$ is the $s$-wave scattering length.
The condensate wave function is given by $\psi(r) = \sqrt{\rho(r)}
e^{i \theta(r)}$, where $\rho(r)$ is the local density of the bosonic
matter in the trap. Long-range ordering in the phase variable
$\theta(r)$ signals the onset of superfluidity or Bose--Einstein
condensation. It will be important for our later discussion that,
consequently and conversely, the {\it destruction} of the
Bose--Einstein condensate proceeds via phase-disordering of the system
through large phase fluctuations $\nabla \theta$. Here, ``large" means phase
fluctuations that give rise to vortices due to the compact nature of
the phase variable $\theta \in [0,2 \pi \rangle$, i.e. phase
fluctuations that have the property $\nabla \times (\nabla \theta) = 2
\pi \ve{n}$, where $\ve{n}$ is an integer-valued vector. These are the
{\it transverse} phase fluctuations, as opposed to the {\it
  longitudinal} ones, which correspond to spin-waves in an $XY$
ferromagnet.
\par 
The potential $V(r)$ is a trapping potential which confines the
Bose--Einstein condensate in a finite region in space.  Moreover, it
is also seen to appear as a spatially varying chemical potential for
the condensate density and as such will set the overall density
profile of the condensate. Increasing $V(r)$ will suppress the
condensate density and vice versa. Furthermore, $\Omega$ is the
angular frequency associated with the rotation of the condensate when
it is spun up, and $\hat{L}_z$ is the corresponding total angular
momentum operator. Finally, the quartic term is a repulsive contact
interaction between the bosons of the condensate which will render the
spectrum of the theory bounded from below. Such an energy functional
is applicable as a coarse grained description of an uncharged phase
coherent condensate \cite{Leggett_review}.
\par
If we formally introduce a vector potential $\ve{A} = (M/\hbar)
(\ve{\Omega} \times \ve{r})$ with $\ve{\Omega} = \Omega \hat{\ve{z}}$,
the energy functional in Eq. (\ref{eq:gpenergy}) can be written
\begin{equation}
  \label{eq:gpenergyvectorpot}
  \begin{split}
    E' = \int dV \Bigl[& -\frac{\hbar^2}{2 M} \psi^{\ast} (\nabla
    -\im \ve{A})^2 \psi   \\
    & + \Bigl( V(r) - \frac{1}{2} M \Omega^2 r^2_{\perp}\Bigr)|\psi|^2
    + \frac{1}{2} g |\psi|^4 \Bigr],
  \end{split}
\end{equation}
where $r^2_{\perp} = x^2 + y^2$. This is formally similar to the
Ginzburg--Landau free energy of a superconductor, apart from the
position-dependent term involving the trapping potential $V(r)$ and
rotation $\Omega$. However, in a superconductor the vector potential
$\ve{A}$ has dynamics and is related to the magnetic field $\ve{B}$
by $\ve{B} = \nabla \times \ve{A}$, and thus we would normally include
a term $(1/2 \mu_0) B^2$ in the free energy density. For extreme
type-II superconductors where the typical penetration depth
$\lambda_L$ of a static magnetic field is much larger than the
coherence length $\xi$, fluctuations in the magnetic field can in many
cases be ignored and this term in the energy can be dropped. The
similarity of the superconductor Ginzburg--Landau theory to that of a
rotating Bose--Einstein condensate is then striking.
\par
So far, the energy functional of Eq. (\ref{eq:gpenergyvectorpot}) is a
continuum theory, but for the purposes of carrying out computer computations, 
it is more convenient to discretize space into a lattice and to rescale the wave
function to avoid prefactors. That is, we let the wave function
$\psi(\ve{r}) \rightarrow (\sqrt{M}/\hbar) \psi_i$, so that it is only
defined on vertices $i=1,\ldots,L^3$, separated by a lattice constant
$a$.  We also replace the gradient term with a gauge invariant lattice
difference,
\begin{equation}
  \label{eq:latticereg}
  \psi^{\ast} (\nabla -\im \ve{A})^2 \psi \rightarrow \sum_{\mu} 
  | \psi_{i+a\mu} \ex{-\im A_{i\mu}} - \psi_i|^2.
\end{equation}
$i+a\mu$ is the lattice site situated next to site $i$ in direction
$\mu$, and the gauge field $A_{i\mu}$ here lives on the links of the
lattice and it is given by the line integral
\begin{equation}
  \label{eq:gaugelineint}
  A_{i\mu} = \int_{i}^{i+a\mu} \du l A_{\mu}.
\end{equation}
The continuum theory is recovered if we let $a\rightarrow 0$.
\par
For high-$T_C$ superconductors, where $\lambda_L \gg \xi$, it is a
well established approximation only to consider fluctuations in the
phase of the order parameter. This is frequently referred to as the
London approximation\cite{Brandt_Review}, in which we simply assume a
condensate of Cooper pairs to exist by having a finite and fixed
$|\psi_i| = |\psi|$, since in the end, it is not the depletion of the
number of Cooper pairs that is responsible for destroying
superconductivity, but rather the proliferation of vortex loops
originating with violent transverse phase fluctuations in the
superconducting order parameter
\cite{Tesanovic1999,NguyenSudbo1999}.  High-$T_C$ superconductors are
extreme type-II superconductors, which in a certain sense means that
the (renormalized) charge of the condensate is small
\cite{Tesanovic1999,NguyenSudbo1999}.  This suppresses gauge field
fluctuations and an external magnetic field therefore only acts as a
frustration on the system via minimal cupling to a fixed external
vector potential, just as in the above
Eq. (\ref{eq:gpenergyvectorpot}). Superfluids and Bose--Einstein
condensates have zero charge and may in this sense be viewed as the
ultimate extreme type-II superconductors where all vestiges of the
fluctuating gauge field in the problem have vanished. In these systems
we therefore expect that the phase only approximation is essentially
exact. Increasing the temperature, large fluctuations in the phase
$\theta$ makes the condensate incoherent and non-superconducting
before $|\psi|$ vanishes, see for instance Fig. 1 of
Ref. \onlinecite{NguyenSudbo1999}.  In fermionic pair condensates
\cite{Greiner1,Greiner2}, it has been strikingly demonstrated that one
may have pairing without superfluidity
\cite{Zwierlein1,Zwierlein2,Zwierlein3,Schunck2007}, providing further
confirmation of the above ideas in a completely different setting than
extreme type-II superconductors or $^4$He\cite{Onsager_vortexloop}.

The right hand side of Eq. (\ref{eq:latticereg}) can be rewritten
\begin{equation}
  \label{eq:londonapprox}
  | \psi_{i+a\mu} \ex{-\im A_{i\mu} - \psi_i}|^2 = |\psi|^2 [2 - 2
  \cos (\Delta_{\mu} \theta_i - A_{i\mu})],
\end{equation}
using the lattice difference operator $\Delta_{\mu} \theta_i
=\theta_{i+a\mu} - \theta_i$. A slight generalization is instead to
replace $\psi_{i+a\mu}$ and $\psi_i$ in Eq. (\ref{eq:latticereg}) with
their average as this will allow for a \emph{non-uniform condensate
  density}, which is what we want to study here.
\par
Since we will only include phase fluctuations in the Monte Carlo
computations, any terms {\it not} containing $\theta$ will represent
mere constant shifts in the total energy and we will consequently drop
them. We thus arrive at a simple, effective energy
\begin{equation}
  \label{eq:nonuniformxy}
  E = \sum_{i \mu} P_{i \mu} \cos(\Delta_{\mu}\theta_i -
  A_{i \mu}).
\end{equation}
The sum is over all positions and directions $x, y, z$, and except for
a factor $M/\hbar^2$, the position dependent coupling $P_{i \mu}$ is
nothing but the condensate density at the link from site $i$ in
$\mu$-direction. If this factor is set to unity, Eq.  (\ref{eq:nonuniformxy}) 
reduces to the well known uniformly frustrated 3D $XY$ model, used for modelling
the melting of the vortex-line lattice in  uniform bosonic condensates and extreme 
type-II superconductors (see e.g.  Refs. \onlinecite{3DXY1,3DXY2,3DXY3,3DXY4,3DXY5}). 
The local vorticity can be calculated from a phase configuration by summing the 
gauge-invariant phase difference around each plaquette in the numerical grid,
\begin{equation}
  \label{eq:vorticity}
  \sum_{\square_j} (\Delta_{\mu} \theta_i - A_{i\mu}) = 2 \pi n_j.
\end{equation}
Here, $n_j$ is the number of vortices penetrating a plaquette
$\square_j$ in the positive direction. The spatial (radial) variation
of $P_{i\mu}$ reflects directly the spatial variation {\it in the
  ground state} of the system of the quantity $|\psi(r)|^2$ due to the
spatial variation of the effective chemical potential $V(r) -
\frac{1}{2} M \Omega^2 r^2_{\perp}$ appearing in
Eq. (\ref{eq:gpenergyvectorpot}).
\par
We mention in passing that {\it{longitudinal}} phase fluctuations are
innocuous in $3$D, and hence need not be considered for the purposes
of studying phase transitions in the system. Such phase-fluctuations
are incapable of destroying the superfluid density in $3$D. In
$2$D they suffice to render the system {\it critical} at any finite
temperature below the Kosterlitz--Thouless transition, meaning that
phase-correlations 
$G({\bf r}-{\bf r}') \equiv \langle e^{i \theta({\bf r})} e^{- i \theta({\bf r}')} \rangle$ 
exhibit a quasi-long range {\it power-law} decay $G(r) \sim 1/r^\eta$ with a temperature-dependent
exponent $\eta$. In neither case
are longitudinal phase fluctuations capable of driving the system {\it through} a phase transition
and into a phase with short-range, {\it exponentially} decaying phase-correlations
$G(r) \sim e^{-r/\xi}$, where $\xi$ is the correlation length.
\par
It should be noted that the model Eq. (\ref{eq:nonuniformxy}) does not
apply to the Lowest Landau Level regime (LLL) of an atomic
Bose--Einstein condensate. The latter can be identified through the
ratio of the interaction energy scale to the level spacing of the
transverse harmonic confinement in an axially symmetric trap,
\begin{equation}
  \label{eq:LLLparameter}
  \lambda = \frac{4 \pi \hbar^2 a_s n }{M \hbar \omega_{\perp}},
\end{equation}
where $n$ is the particle number density, $a_s$ is the s-wave
scattering length, $M$ is the particle mass and $\omega_{\perp}$ is
the trap frequency. The LLL approximation is considered to be valid
when $\lambda \ll 1$ \cite{TF2,TF3}. However, we expect Eq.
(\ref{eq:nonuniformxy}) to be adequate under the conditions of those
experiments which meet the following naive requirement. The
intervortex distance $2r_0 = 2 \sqrt{\hbar / M \Omega}$ should be
substantially larger than the healing (or coherence) length $\xi =
\hbar / M \omega_{\perp} R_{\perp}$, where $R_{\perp}$ is the radius
of the system perpendicular to the axis of rotation
\cite{Madison,Coddington1,Schweikhard,Smith,Coddington2,Muniz}.

\section{\label{sec:montecarlo} MONTE CARLO COMPUTATIONS}
The Monte Carlo computations are performed with the standard
Metropolis-Hastings algorithm \cite{metropolis,hastings}, and the
updates are always local.  Initially, all phases are chosen equal to
zero, but any other configuration would suffice, provided that the
system is allowed to thermalize for a sufficiently long time. We then
proceed systematically through all lattice sites one by one and
propose trial updates of the local phases $\theta_i \rightarrow
\theta_i + \delta \theta_i$, where $\delta \theta_i$ is drawn from a
uniform distribution $[-\pi, \pi \rangle$.  Each trial update is accepted
with a probability $p$ determined from the energy difference $\Delta
E$ of the configurations before and after the update,
\begin{equation}
  \label{eq:acceptprobability}
  p =
  \begin{cases}
    1 & \text{if $\Delta E < 0$},\\
    \ex{-\Delta E / T} & \text{if $\Delta E \geq 0$}.
  \end{cases}
\end{equation}
The result is a Markov chain of configurations that can be used to
estimate the partition function
\begin{equation}
  \label{eq:partitionfunc}
  Z = \sum_{\{ \theta \}}\ex{-E / T },
\end{equation}
where the sum is over all possible sets of the phase. An obvious
effect of generating phase configurations according to the Boltzmann
distribution is how the amount of fluctuations is controlled by the
temperature $T$. At high temperatures, the phases will fluctuate more
easily whereas they tend to freeze when $T$ is lowered.  In this
paper, we will use units of temperature that are such that the
critical temperature of the model in Eq. (\ref{eq:nonuniformxy}), with
$P_{i\mu} = 1$ and $A_{i\mu} = 0,$ is $T_C \simeq 2.2$. By inspecting
the energy Eq.  (\ref{eq:nonuniformxy}) we see that in a Monte Carlo
computation, the density profile $P_{i\mu}$ effectively works as the
inverse of a position dependent effective temperature, such that
\begin{eqnarray}
  \label{eq:effectiveT}
  T_{\rm{eff}} = \frac{T}{P_{i\mu}},
\end{eqnarray}
whence we expect phase fluctuations to depend strongly on the density
profile of the condensate. This is an exact statement within the
phase-only approximation. In particular, this means that the trapped
condensates effectively are ``warm'' (in the sense of being close to
the condensation temperature) wherever the ground state density is
low. In regions of low density we therefore expect more violent vortex
fluctuations. As will be shown below, this is typically the case close
to the edge of the trap, but may also be true close to the center of
the trap for anharmonic trapping potentials with a dip in the
condensate density at the center of the trap.
\par
We define the process of going over all sites once as one sweep and
measure the Monte Carlo time in units of the sweeps. Initially at each
temperature, all realizations of the system are thermalized with at
least 100 000 sweeps to make sure they fluctuate around the
equilibrium state before any measurements are made. To calculate
thermal averages, we sample the configuration every 100th sweep. The
numerical grid is cubic with a linear extension $L = 72$ (for the case
of the uniform cylinder, we also consider $L= 36$).

\subsection{\label{sec:helicity} The helicity  modulus}
Phase coherence in a vortex system is probed by computing the helicity
modulus $\Upsilon_{\mu}$, equivalently the \emph{superfluid density},
defined as the change in free energy $F = - T \ln Z$ as a result of a
phase twist $\tilde{\ve{\Delta}}$ applied in the the following way,
\begin{equation}
  \label{eq:energywithtwist}  
  E[\tilde{\ve{\Delta}} ] = \sum_{i \mu} P_{i \mu} \cos(\Delta_{\mu}\theta_i -
  A_{i \mu} - \tilde{\Delta}_{\mu}).
\end{equation}
The expansion of the free energy is even in $\ve{\Delta}$, which can
be viewed as a change in the boundary conditions of the system. For
small deviations from periodic boundary conditions, the leading
behaviour is quadratic and the helicity modulus is given as the
coefficient to the second order term,
\begin{eqnarray}
  \label{eq:helicity}
  \Upsilon_{\mu} & \equiv &  
  \frac{1}{L^3}~\frac{\delta^2 F}{\delta \tilde{\Delta}_\mu^2} \bigg|_{\tilde{\Delta}_\mu=0} \nonumber \\
  & = & \frac{1}{L^3} \left\langle \sum_{i} P_{i
      \mu} \cos (\Delta_{\mu}\theta_i-A_{i\mu}) \right\rangle
  \nonumber \\
  & - & \frac{1}{T L^3} \left\langle \left[\sum_{i} P_{i \mu} \sin
      (\Delta_{\mu}\theta_i-A_{i\mu})\right]^2 \right\rangle.
\end{eqnarray}
The twist can be applied in any direction, and without rotation the
response in terms of $\Upsilon_{\mu}$ is equal for all $\mu =
x,y,z$. When the temperature is increased from low to high, thermal
fluctuations gradually destroy phase coherence. Consequently, the
renormalized superfluid density continuously evolves from a finite
value to zero at some critical temperature.  On the other hand, in a
rotating system the helicity modulus will be different for the $x$-
and $y$-directions than along the axis of rotation. Both cases are
however important since $\Upsilon_{x}$ and $\Upsilon_{y}$ carries
information on numerical pinning. In the computations, we choose the
vector potential so that $\nabla \times \ve{A} = (0,0,2 \pi f)$, where
$f$ is the number of rotation-induced vortices per numerical
grid-plaquette in the $xy$-plane. For technical reasons, we restrict
the filling fraction $f \leq k / L^2$ with $k = 1,2,3,\ldots$, but at
the same time the density of vortex lines should not be too high in
order to avoid artificial pinning to the underlying numerical grid
\cite{footnote_gridpinning}. This can be probed by the helicity
modulus in the transverse directions.  A zero value of $\Upsilon_x$
and $\Upsilon_y$ means that the vortex line lattice is free to move
translationally with respect to the grid. For very low temperatures,
there will always be pinning, but the pinning should disappear well
below the temperature at which the lattice melts. This melting on the
other hand, is characterized by a discontinuous jump in the superfluid
density measured parallel to the vortex lines, i.e. a jump in
$\Upsilon_z$ to zero.
\par
In non-uniform systems, we encounter a problem with the above
definition of the helicity modulus, since $P_{i\mu}$ equal to or close
to zero at the system edges will favour fluctuations at all
temperatures in these regions. The global $\Upsilon_{\mu}$ mixes
together information on the amount of fluctuations in all regions, and
the interpretation is therefore less useful. To obviate this
difficulty, we introduce a modified helicity modulus in $z$-direction,
defined in a selected region between two cylinders of radii $R_1$ and
$R_2$. We do so by applying a twist
\begin{equation}
  \label{eq:twist}
  \tilde{\ve{\Delta}}(r_{i z}) = 
  \begin{cases}
    \tilde{\Delta} \ve{\hat{z}} & \text{if $R_1 \le r_{i z} < R_2$},\\
    0 & \text{otherwise},
  \end{cases}
\end{equation}
and defining the modified helicity modulus as follows,
\begin{align}
  \label{eq:cylheli}
  \tilde{\Upsilon}_z(R_1,R_2&) \equiv \frac{1}{N^{\prime}}
  \left\langle {\sum}^{\prime} P_{iz} \cos(\Delta_{z} \theta_i -
    A_{i z})\right\rangle \nonumber \\
  &- \frac{1}{T N^{\prime}} \left\langle \left[ {\sum}^{\prime} P_{i
        z} \sin(\Delta_{z}\theta_i - A_{i z})\right]^2\right\rangle.
\end{align}
Here, $\sum^{\prime}$ is over all sites where $\tilde{\ve{\Delta}}(r_{i z})$
is nonzero (depending on $R_1$ and $R_2$) and $N^{\prime}$ is the
number of these sites. The local $\tilde{\Upsilon}_z(R_1,R_2)$
provides a measure of the phase coherence more locally than the global
quantity, and will therefore be used to investigate the character of
the vortex state at different positions in a trapped Bose--Einstein
condensate.

\section{\label{sec:uniform} UNIFORM SYSTEMS}
As a warmup to the results to be presented below, we first consider
two cases of uniform systems, namely the infinite uniform system and a
cylindrical confinement geometry with uniform density.  The former in
particular allows connections to be made to the vast literature on
vortex physics of extreme type-II superconductors such as high-$T_C$
superconductors.
 
\subsection{\label{sec:infinite} Infinite uniform system}
\begin{figure}[t]
  \vspace{.5cm}
  \includegraphics[width=.5\textwidth]{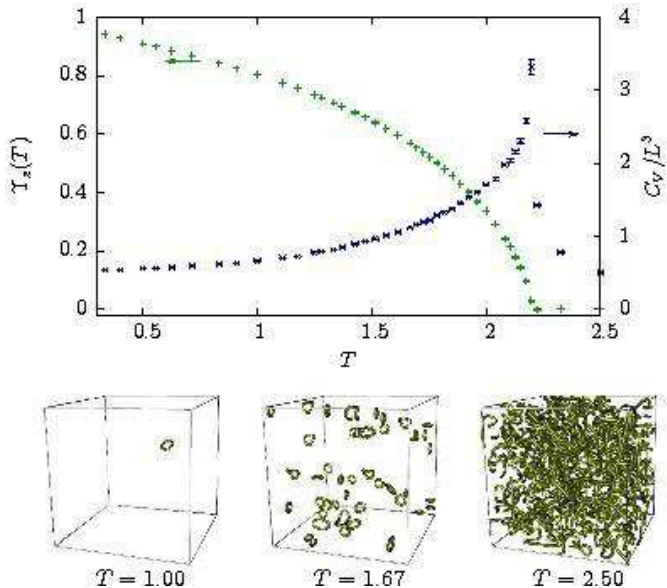}
  \caption{(Color online) Monte Carlo results for a uniform system
    with periodic boundary conditions. The helicity modulus
    (superfluid density) $\Upsilon_z$ along the $z$-direction is
    plotted ($+$), but $\Upsilon_x$ and $\Upsilon_y$ give equal
    results. The specific heat $C_V/L^3$ ($\times$) has a peak at
    $T_C$, where $\Upsilon_z$ vanishes. In the second row, random
    snapshots of the vortex configurations are shown for three
    temperatures. These are sections with $16^3$ lattice sites cut out
    from the $72^3$ system.}
  \label{fig:UniformNoRot}
\end{figure}

We begin with a review of the Monte Carlo results for a uniform
system, i.e. with $P_{i\mu} \equiv 1$ in Eq. (\ref{eq:nonuniformxy}),
and we first consider the non-rotating case. This is the standard
uniform 3D $XY$ model, which has been studied extensively elsewhere,
see for example Ref.  \onlinecite{3DXY1,3DXY2,3DXY3,3DXY4,3DXY5}. At
low temperatures, the phases tend to align as spins in a ferromagnet,
and the system is frozen in a stiff state where any change in boundary
conditions is associated with a large response in free
energy. Consequently, the helicity modulus is close to
unity. Equivalently, the superfluid density is close to the ground
state density $P_{i \mu}$. In Fig. \ref{fig:UniformNoRot} is shown the
helicity modulus $\Upsilon_z$ along the $z$-direction, but
$\Upsilon_x$ or $\Upsilon_y$ would give similar results. As the
temperature is increased, the relevant phase fluctuations start to
appear as vortex loops, in numbers that gradually increase with
temperature. This reduces the superfluid density. However, not until
the vortex loops loose their line tension (free energy per unit
length) at the critical temperature $T_C \simeq 2.2$, will
$\Upsilon_z$ vanish completely.
The phase transition is continuous and accompanied by a diverging
length scale in the thermodynamic limit and a developing singularity
in the specific heat $C_V$ as shown in Fig. \ref{fig:UniformNoRot}.
High precision measurements of the critical exponents can be found in
Ref. \onlinecite{Hasenbusch}

In the second row of Fig. \ref{fig:UniformNoRot}, we show some
snapshots from the computations where the vortices have been calculated
via Eq. (\ref{eq:vorticity}) and plotted in a 3D volume. The vortex
radius is \emph{chosen} to be 0.4 times the grid spacing for a
convenient visualization, and this should not be associated with the
true core size. Additionally, the vortices (especially the vortex
\emph{lines} in the rotating system we present below) exhibit sharp
bends at short length scales. These bends result from the numerical
grid. In fact, the vortices shown in the images are splines; the
precise form of the vortices in the computations is straight line
segments connected in perpendicular corners. Nevertheless, the model
has proved to be accurate for vortex fluctuations at scales larger
than the grid spacing \cite{3DXY1,3DXY2,3DXY3,3DXY4,3DXY5}. Hence, the
3D snapshots provide useful hints to the state the system is in at
different temperatures. Only occasional and small vortex loops appear
when the temperature is low, but eventually they fill the whole
volume. Above $T_C$, the system is in a state dominated by a tangle of
tensionless vortex loops of all sizes, and there is no phase coherence
or superfluidity.
\par
\begin{figure*}[htbp]%%[htbp]
  \includegraphics[width=.83\textwidth]{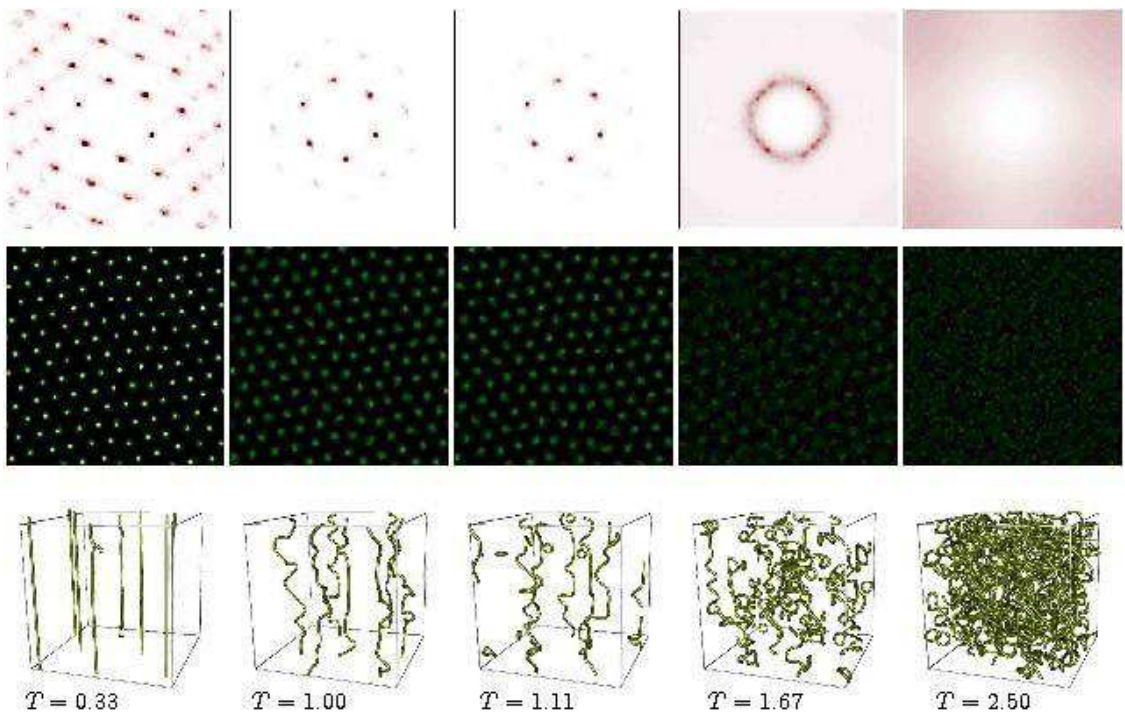}
  \caption{(Color online) For each temperature, given below the
    columns, the structure function $S (\ve{k_{\perp}})$ is shown in
    the upper row ($S(0)$ is removed, and in the two rightmost images 
    the colorscale is magnified by a factor 25). There are sharp peaks for 
    the characteristic Bragg vectors at the lowest temperatures before a
    transition to a ring structure corresponding to a vortex liquid
    phase at $T=1.67$. In the second row we show the real-space
    equivalent to the structure function, the $xy$-positions of the
    vortices, integrated over $z$-direction. The averages of the first
    two rows are calculated from 100 000 Monte Carlo sweeps. In the
    third row, we show sections of size $16^3$ from snapshots of the
    vortex configurations.}
  \label{fig:UniformStructure}
\end{figure*}

In a rotating system, the scenario is different. The ground state is
the famous Abrikosov lattice \cite{Abrikosov,baym}, where
straight rotation-induced vortex lines arrange themselves in a
triangular pattern, though with some defects due to the
incommensurable underlying square numerical grid. Here, we present
computation results from a system with filling fraction $f = 1/36$, and
the structure of the vortex line lattice can be seen in
Fig. \ref{fig:UniformStructure}.
In the upper row is shown the structure function,
\begin{equation}
  \label{eq:structfunc}
  S (\ve{k_{\perp}}) = \frac{1}{f L^3} \left\langle \left|
      \sum_{\ve{r_{\perp}}} n(\ve{r_{\perp}})
      \ex{\im \ve{k_{\perp}\cdot \ve{r_{\perp}}}}\right|^2
  \right\rangle,  
\end{equation}
where $ n(\ve{r_{\perp}})$ is the local vorticity given by
Eq. (\ref{eq:vorticity}) and the sum runs over the positions of all
plaquettes in the $xy$-planes. The $\ve{r_{\perp}}$- and
$\ve{k_{\perp}}$-vectors are perpendicular to the axis of
rotation. The structure function exhibits sharp peaks for the
characteristic Bragg vectors of the vortex line lattice when the
temperature is low and the system close to its ground state. As the
effect of thermal fluctuations sets in and the vortex lines gets more
prone to bending, the Bragg-peaks are weakened, and eventually there
is a transition from a sixfold-symmetric structure to a ring structure
in $S (\ve{k_{\perp}})$. This is the hallmark of a vortex liquid phase
where the vortices still posess a line tension which disappears in a
crossover transition at an even higher temperature. The second row in
Fig. \ref{fig:UniformStructure} is the real-space equivalent of the
above, where the local vorticity $n(\ve{r_{\perp}})$ is integrated
along the $z$-direction so that closed vortex loops will be cancelled
or averaged out. Both the Fourier- and real-space versions of the
structure function are thermal averages calculated from 100 000 Monte
Carlo sweeps. In the left panels of the second row, bright spots
correspond to straight lines in relatively stable positions. Higher
temperatures result in increased bending of the lines, and the bright
spots develop into smeared-out regions until it is no longer possible
to distinguish individual vortex lines in the rightmost panel. Insight
into the bending mechanism can be obtained from the 3D sections in the
third row, which are snapshots corresponding to those of the
non-rotating case in Fig. \ref{fig:UniformNoRot}.  In principle, such
snapshots from a Monte Carlo computation could produce \emph{any}
possible configuration, but a state far from equilibrium is highly
unlikely. We therefore assume the pictures to be representative and
useful indications of the system's state at a given
temperature. Compared to the non-rotating system, we see that there is
much going on in terms of vortex fluctuations even at low temperatures
like $T = 1.00$, in the units we defined below
Eq. (\ref{eq:partitionfunc}). The energy cost associated with
elementary vortex excitations, i.e. a closed vortex-loop, is less when
there are vortex lines already present. A loop and a line can merge to
produce a bend in the vortex line and enough bends will result in a
melting transition at $T_M$.  This is succeeded by a vortex loop
blowout inside the vortex-liquid phase similar to what happens in the
non-rotating case.

\begin{figure}[htbp]
  \includegraphics[width=.5\textwidth]{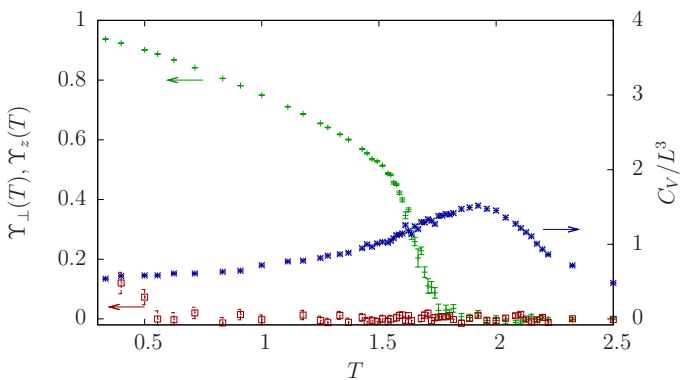}
  \caption{(Color online) The helicity modulus $\Upsilon_z$ ($+$)
    along the axis of rotation is cut off and vanishes when the vortex
    lattice melts. In the transverse direction, a zero
    $\Upsilon_{\perp} = \Upsilon_x, \Upsilon_y$ ($\square$) indicates
    that there is no pinning of the vortices to the numerical
    grid. The rounded peak in $C_V/L^3$ ($\times$) is a remnant of the
    vortex-loop blowout transition in the non-rotating system.}
  \label{fig:UniformHeli}
\end{figure}

The qualitative picture above is supported by the qantitative
measurements presented in Fig. \ref{fig:UniformHeli}.
The vortex loop blowout no longer corresponds to a phase transition,
but a remnant crossover is still indicated by a peak in the specific
heat at a temperature $T > T_M$. A finite size scaling analysis would
however reveal that there is no criticality associated with this broad
peak \cite{NguyenSudbo1999}. On the other hand, the melting of the
vortex line lattice is a first order phase transition characterized by
a discontinuity in the helicity modulus (or superfluid density)
$\Upsilon_z$ along the axis of rotation in the thermodynamic limit. In
a finite system like the one we have simulated, the drop to zero is
continuous, but compared to a non-rotating system, it is much steeper
and takes place at a lower temperature. Finally, note that the
helicity modulus in the transverse direction $\Upsilon_{\perp}$ is
zero for temperatures well below the melting transition, indicating
that the vortex line lattice is not pinned to the numerical grid.

\subsection{\label{sec:cylinder} Cylindrical container}
\begin{figure}[htbp]
%%  \vspace{.5cm}
  \centerline{\hbox{\includegraphics[width=85mm]{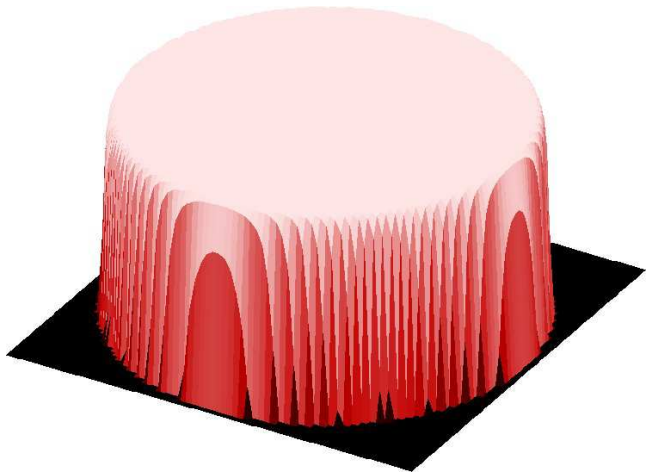}}}
  \caption{(Color online) Radial density profile $P_{i\mu}$ for the
    case of a uniform sylinder. The density is uniform in the
    $z$-direction.}
    \label{fig:uniformcylinder}
\end{figure}

\begin{figure}[htbp]
%%  \vspace{.5cm}
  \centerline{\hbox{\includegraphics[width=85mm]{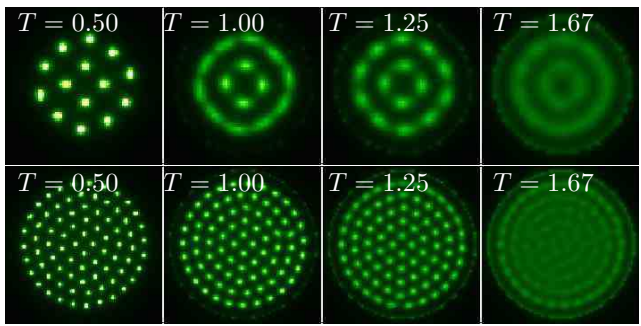}}}
  \caption{(Color online) $xy$ positions of vortices in a cylindrical
    container integrated over $z$-direction, and averaged over every
    tenth of a total of $5 \cdot 10^5$ MC sweeps. Top and bottom rows
    have $L = 36$ and $L=72$, respectively. At $T=0.5, 1.0$, we
    discern circular ordering close to the cylinder wall combined with
    a hexagonally ordered state closer to the center. At $T=1.25,
    1.67$ we observe dominance of angular fluctuations closest to the
    edge. }
    \label{fig:hardboundary}
\end{figure}

\begin{figure}[htbp]
  \centerline{\hbox{\includegraphics[width=85mm]{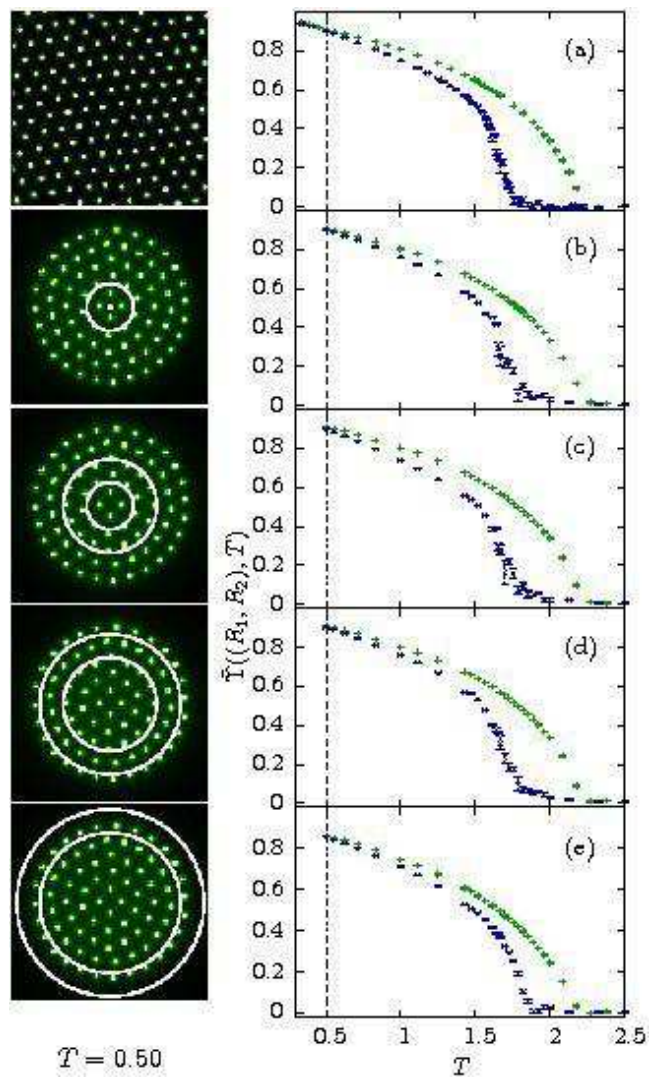}}}
  \caption{(Color online) Results for $\tilde{\Upsilon}_z(R_1,R_2)$ in
    a cylindrical container.  In panels (b)--(e) are shown the
    helicity moduli $\tilde{\Upsilon}_z(R_1,R_2)$ for radii $R_1,R_2 =
    0, R/4,2R/4,3R/4,R$, as indicated by the white circles in the
    images on the left, and compared to the results for the extended
    uniform system, panel (a). The upper curve ($+$) is the helicity
    modulus without rotation, while the lower curve ($\times$) is the
    results with rotation-induced vortices present (filling fraction
    $f = 1/36$). All regions in the cylindrical container behave
    almost as the uniform system.}
      \label{fig:helimodcylinder}
\end{figure}

We next consider the case of a uniform cylinder. This gives a ground
state density profile $P_{i\mu}$ illustrated in
Fig. \ref{fig:uniformcylinder}.
Figs. \ref{fig:hardboundary} and \ref{fig:helimodcylinder} show the
results from computations of vortex matter in a cylindrical container
with such a density profile, given by
\begin{equation}
  \label{eq:cyldensity}
  P_{i\mu}= \Theta(r_{i\mu} - R),
  \end{equation}
  where $\Theta$ is the Heaviside step function $\Theta(x) = 0, x <
  0$; $\Theta(x) = 1, x > 0$.  We have used two different system sizes
  $L$, but with the same filling fraction $f = 1/36$ to see how the radius $R
  = L/2-2$ affects the ordering of the vortices. At low temperatures,
  the computations reproduce orderings with circular distortions of the
  vortex line lattice near the container wall, as predicted for
  ${}^4$He in a zero-temperature treatment of the problem
  \cite{zipf}. For a large number of vortices the system reacquires
  the hexagonal lattice symmetry away from the wall, see
  Fig. \ref{fig:hardboundary} (bottom row).
Increasing the temperature in the case of small number of vortices
(top row of Fig. \ref{fig:hardboundary}) the dominant vortex
fluctuations are associated with angular displacements, while radially
the vortex density remains ordered. For the largest system, with many
vortices, (bottom row of Fig. \ref{fig:hardboundary}) we find
dominance of angular fluctuations only for the vortices situated close
to container wall, while the center of the system does not display
this phenomenon.
\par
The crossover to a uniformly molten vortex system occurs in both cases
only at a higher temperature. The two-step thermal crossover in the
vortex pattern we find is analogous to that in two dimensions where
vortices are point-like objects (see e.g. Ref. \onlinecite{2dm}). There is,
however, a principal difference in our case, since in three dimensions
the vortex line lattice melting is accompanied by significant vortex
bending fluctuations.

Inspecting the modified helicity modulus $\tilde{\Upsilon}_z(R_1,R_2)$
(Fig. \ref{fig:helimodcylinder}) for different regions $(R_1,R_2)$
inside the cylinder, and comparing these results to those of a uniform
system (panel (a)), we find only small differences. The cylindrical
system behaves as the uniform one, and the circular distortions do not
seem to affect the superfluid density. One could argue that the drop
to zero is a little more rounded in the rotating non-uniform system,
but this can be explained by the fact that
$\tilde{\Upsilon}_z(R_1,R_2)$ is calculated for smaller subsystems and
that the finite size effects necessarily should be more severe
here. In fact the cylindrical container is just a uniform system with
different boundary conditions, and in a sufficiently large system the
boundary effects are irrelevant.

\section{\label{sec:nonuniform} NON-UNIFORM SYSTEMS}
In this section we present Monte Carlo results from systems with
non-uniform density profiles $P_{i\mu}$. The finite size of the
systems is now a wanted feature and closer to the real situation with
ultra-cold atoms, and we do not need to do finite size
scaling. Strictly speaking, the only possibility for transitions are
of crossover nature. To reduce the surface effects and because we
model elongated systems, we employ periodic boundary conditions in the
$z$-direction. Global quantities such as the ordinary helicity modulus
$\Upsilon_{\mu}$ and the structure function $S (\ve{k_{\perp}})$ have
no rigorous meaning in these systems, and we rather use local versions
like the modified helicity modulus $\tilde{\Upsilon}_z(R_1,R_2)$ and
the local vorticity $n(\ve{r_{\perp}})$.  The 3D snapshots are also
useful indicators on the mechanisms involved.

\subsection{\label{sec:harmonic} Harmonic trap}
\begin{figure}[htbp]
  %% \vspace{.5cm}
  \centerline{\hbox{\includegraphics[width=85mm]{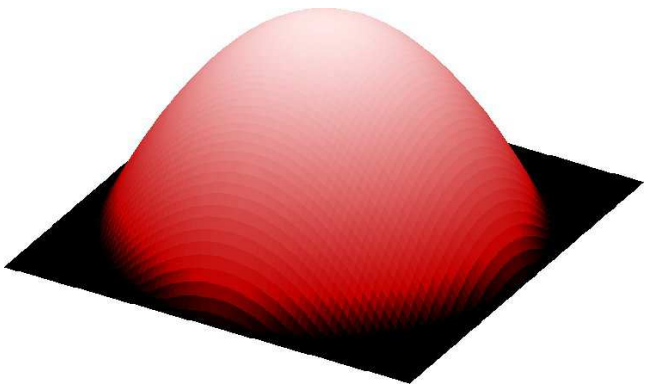}}}
  \caption{(Color online) Radial density profile $P_{i\mu}$ for the case of a
    harmonic trap.}
    \label{fig:harmonictrap}
\end{figure}

For a system in a harmonic trap, we choose a density profile according
to the Thomas--Fermi approximation \cite{TF1,TF2,TF3} with the shape
of an inverse parabola, $P_{i\mu} = P^h \left( \frac{r_{i \mu}}{R
  }\right) ~ \Theta( R - r_{i \mu})$, where $\Theta$ is
the Heaviside step function $\Theta(x) = 0, x < 0$, $\Theta(x) = 1, x
> 0$, and
\begin{equation}
  \label{eq:harmonicdensity}
  P^h (x) = 1- x^2.
\end{equation}
The density gradient in a trap can alternatively be viewed as an
effective temperature gradient in a uniform system, see
Eq. \ref{eq:effectiveT}. It is clear that for low, but finite, actual
temperatures $T$, there will be a finite area near the edge of the
cloud which \emph{effectively} would be at a high enough temperature to
feature an annulus of tension-less tangle of vortices. This is a phase
where the vortex line tension has vanished through the proliferation
of vortex loops. The ground state density profile $P_{i\mu}$ is shown
for the harmonic trap in Fig. \ref{fig:harmonictrap}.
The question is whether there is also a vortex liquid region in
between the ordinary vortex line lattice and the tensionless annulus,
which represents the true boundary of the condensate.
\begin{figure*}[htbp]
  {\includegraphics[width=.83\textwidth]{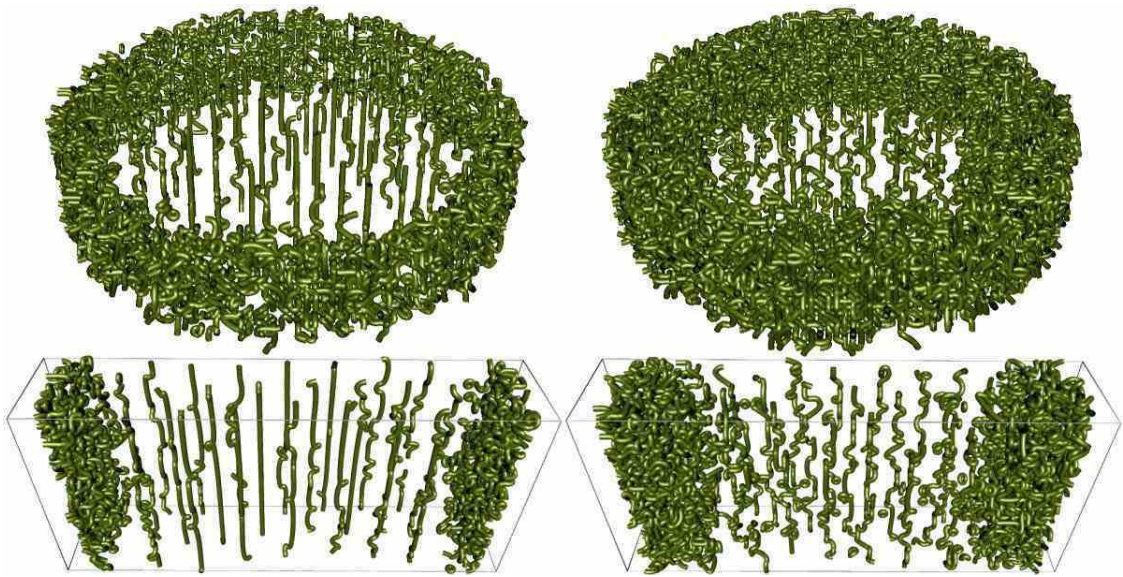}}
  \caption{(Color online) Snapshots of vortex configurations in a
    rotating trapped BEC at $T = 0.5$ (left figures) and at $T=1.0$
    (right figures). The top row shows a selection of $16$ out of $72$
    layers in $z$ direction.  The bottom row shows smaller selections
    in the $xy$ plane, but $32$ out of $72$ layers in $z$ direction.
    Fluctuations are minimal in the trap center, and increase towards
    the edge of the trap. A distinct front separating regions of
    ordered and disordered vortices is easily identified.}
    \label{fig:3dsnapshotharmonic}
\end{figure*}
The 3D snapshots in Fig. \ref{fig:3dsnapshotharmonic} illustrate how
the vortices are stiffer in the central part than further out towards
the edge where there is an annulus with a tangle of tension-less
vortices. At the higher temperature on the right, this region has
grown, but simultaneously the amount of bends in the vortex lines in
the center has increased.
\par
\begin{figure}[htbp]
    \centerline{\hbox{\includegraphics[width=85mm]{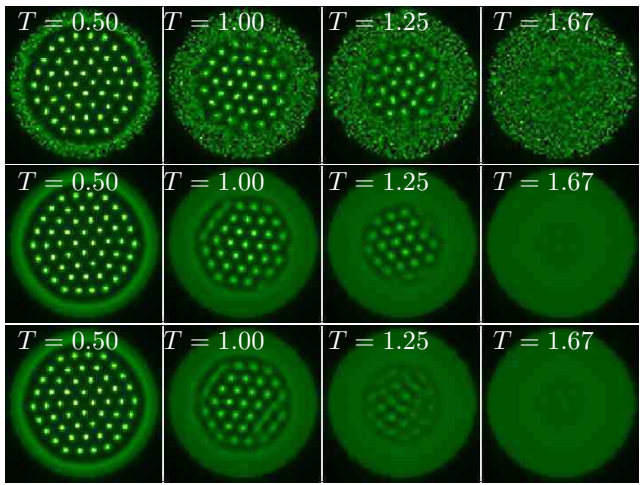}}}
    \caption{(Color online) $xy$-positions of vortices in a trapped
      BEC integrated over $z$-direction. In the top row, we show
      snapshots, while the middle and bottom rows show averages of
      $10^5$ and $5 \cdot 10^5$ MC sweeps, respectively.  Every tenth
      configuration has been sampled. This provides information on the
      stability of the ordered region and the evolution of the
      disordered region as $T$ varies.}
\label{fig:vortexpositionsharmonic}
\end{figure}

Further insight into the stability of the vortex line lattice can be
obtained from the $z$-integrated vorticity in
Fig. \ref{fig:vortexpositionsharmonic}. The top row consists of
snapshots and already here it is easy to separate the ordered lattice
region from the disordered one, since straight vortex lines are seen
as bright spots while bent vortices result in smeared spots or even
smeared regions. This observation may be related to experiments, where
at least for non-equilibrated vortex systems the $z$-integration
renders vortices essentially indistinguishable
\cite{bretin,Abo-Shaer}. We have used a filling fraction $f =
1/36$. Taking parameters from Ref.  \onlinecite{Ketterle}, and using
$\Omega = (h/M) N_v/2 \pi R^2$ where $N_v$ is the number of vortices
in the trap, we find $\Omega \sim 100 \rm{Hz}$. Since $\omega_{\perp}
\sim 500 \rm{Hz}$\cite{Ketterle}, this puts us well outside the LLL
regime.

We find that a well ordered vortex line lattice extends over most of
the system at $T = 0.50$. Note also the absence of circular
distortions for the vortices at the edge of the system, as opposed to
the situation in the cylindrical container. At $T = 1.67$ we can still
distinguish $2-3$ central vortices where the density is the highest in
the snapshot. However, in the thermal averages of the second and third
row these vortices are no longer possible to detect due to their
thermally fluctuating positions.  The thermal averages are created by
averaging snapshots like the ones in the first row over 100 000 Monte
Carlo sweeps in the second row and 500 000 sweeps in the third.
Indeed, in a finite system, the averaging will eventually produce a
complete smearing even in the center of the trap since there is only a
finite energy barrier to translate or rotate even a perfect vortex
line lattice. Signatures of this effect can be seen in the difference
between the $T=1.25$ pictures of the second and third row of
Fig. \ref{fig:vortexpositionsharmonic}. However, it is important to
note that the Monte Carlo time scale of the fluctuations we observe in
the ordered regions, is dramatically larger than those related to the
fluctuations in the disordered ones. Therefore, it does make sense to
speak of ordered and disordered regions in these systems.
%\par
%In Fig. \ref{fig:vortexpositionsharmonic}, 
%the intervortex distance is given by $2 r_0 =
%      2 \sqrt{\hbar/M \Omega}$ and the healing length is given by $\xi = \hbar/\sqrt{2
%        M g_{2D} n} \leq c a/2$, where $a$ is lattice constant and $c$
%      is a constant of order, but less than, unity. Hence, $2 \hbar
%      \Omega/g_{2D} n \leq c^2/9$, which according to
%      Refs. \onlinecite{Coddington1,Coddington2} validates our
%      approach of using the London approximation as opposed to the LLL approximation. 

\begin{figure}[htbp]
  \centerline{\hbox{\includegraphics[width=85mm]{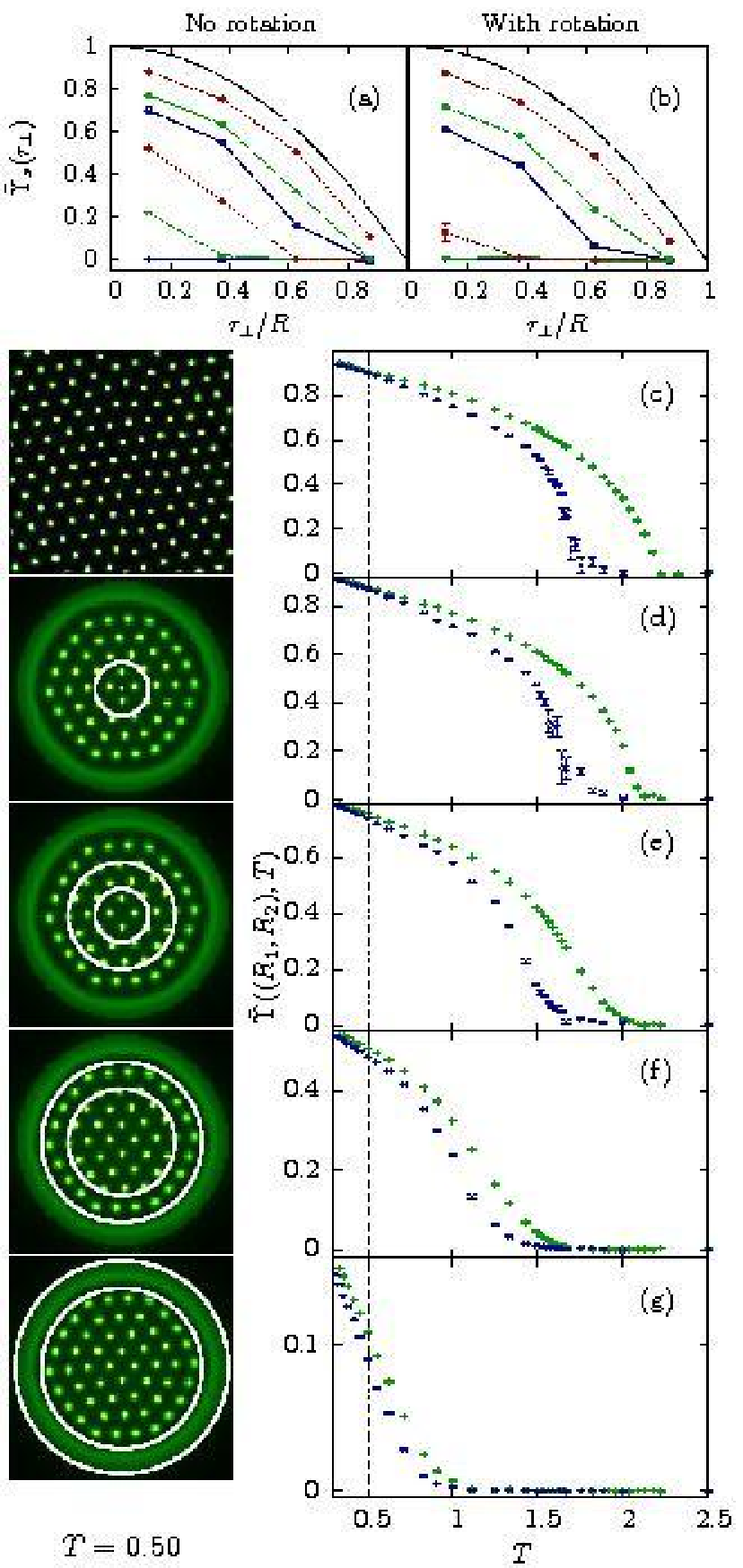}}}
  \caption{(Color online) Results for $\tilde{\Upsilon}_z(R_1,R_2)$.
    The two top panels show computation results for the thermal
    depletion of the superfluid density in a harmonically trapped
    condensate ($r_{\perp}$ is the distance from the center of the trap) at
    the temperatures $T=2.50$ (the lowermost curve), $T=2.00$,
    $T=1.67$, $T=1.25$, $T=1.00$, $T=0.50$.  The uppermost curve is
    the pure ground state density profile $P_{i \mu}$. In panels
    (c)--(g), the upper curve ($+$) is the helicity modulus without
    rotation, while the lower curve ($\times$) the helicity modulus
    with rotation-induced vortices with filling fraction $f = 1/36$ as
    functions of temperature. Panel (c) shows $\Upsilon_z$ for a cubic
    uniform system with periodic boundary conditions.  The remaining
    panels (d)-(g) show $\tilde{\Upsilon}_z(0,R/4)$,
    $\tilde{\Upsilon}_z(R/4,2R/4)$, $\tilde{\Upsilon}_z(2R/4,3R/4)$,
    and $\tilde{\Upsilon}_z(3R/4,R)$, respectively. The vortex plots
    on the left (obtained at $T=0.50$) defines the radii $R_1$ and
    $R_2$ as white circles.}
      \label{fig:helimodharmonic}
\end{figure}
We then again turn to the question of the character of the vortex
state in the disordered region and the possibility of a vortex liquid
layer here. For this we use the modified helicity modulus
$\tilde{\Upsilon}_z(R_1,R_2)$ and compare the results to the extended
uniform system as we did for the cylindrical container. The results
are shown in Fig. \ref{fig:helimodharmonic}, where both rotating and
non-rotating systems are considered. If one were to observe no
appreciable difference in the temperature dependence of the helicity
with and without rotation, one would conclude that the demarcation
line seen in the images of Fig. \ref{fig:vortexpositionsharmonic}
separates an ordered region from tension-less vortex tangle, with no
discernible vortex liquid region.
\par
The measurements indeed show that the presence of a rotation
significantly reduces the temperature at which $\tilde{\Upsilon}_z$
vanishes. This reduction relative to the case of no rotation decreases
with increasing $R_1,R_2$. That is, panel (d) is similar to panel (c)
(no trap), whereas in panel (g) there is little difference between
$\tilde{\Upsilon}_z(R_1,R_2)$ with and without rotation.  Thus, for
the latter case the presence of vortices essentially does not
influence $\tilde{\Upsilon}_z$, and the destruction of superfluid
density is driven by the proliferation of vortex loops. In panel (a)
and (b) we have plotted $\tilde{\Upsilon}_z$ as a function of the
distance $r_{\perp} = (R_1 + R_2)/2$ from the center of the trap for
different temperatures. The same conclusions may be drawn here. For
the largest $r_{\perp}$ there are small differences between (a) and
(b), but closer to the center, the helicity modulus suddenly drops to
zero when the temperature is increased in the system with rotation
(b). This corresponds to the melting transition. The reason why we
only find such characteristics in the central part, is that the
density is almost uniform here. Further out, the density gradient is
simply too large and the vortex line lattice crosses directly over to
the tension-less tangle with no visible tension-full vortex liquid
region in between. This is consistent with the experiments showing a
very regular edge structure for systems with large number of vortices
\cite{Madison,Coddington1,Schweikhard,Smith,Coddington2,Muniz,Ketterle}.

\subsection{\label{sec:anharmonic} Anharmonic trap}
Experimentally, the traps that are used to confine the Bose--Einstein
condensates may also be made anharmonic \cite{bretin}. To study the
effect of a harmonic plus quartic trapping potential, we will use a
modified Thomas--Fermi density distribution which varies in the
$xy$-plane as $\alpha_1 + \alpha_2\ve{r_{\perp}}^2 + \alpha_3
\ve{r_{\perp}}^4$. In an experiment with a rotating atomic gas, the
ratio $\alpha_3/\alpha_2$ depends on the rotation frequency, but for
technical reasons we cannot change $\Omega$ during the
computations. Thus, we choose a fixed number of vortices by specifying
$f$ and vary only the temperature during computation runs for fixed
$\alpha_3/\alpha_2$-ratios. To be precise, we have chosen the
following density distribution $P_{i\mu} = P^a \left( \frac{r_{i
      \mu}}{R }\right) ~ \Theta( R - r_{i \mu})$, where $\Theta$ is
the Heaviside step function $\Theta(x) = 0, x < 0$, $\Theta(x) = 1, x
> 0$, and
\begin{equation}
  \label{eq:anharmdensity}
  P^a(x) = 
    \frac{4(1+\alpha)}{4(1+\alpha)+\alpha^2} 
    \left\{ 
    1 + \alpha  x^2 - (1+\alpha) x^4  
    \right\}.
\end{equation}
This ground state density profile has the property that its maximum is always
unity. Consequently, the regions with the highest density should be
comparable to both the uniform system and to the center of the
harmonically trapped system at any given temperature. The value of
$\alpha$ determines the shape of the density profile. Note that
$\alpha=0$ corresponds to a pure quartic trapping potential.  We
present results for this in the upper row of
Fig. \ref{fig:vortexpositionsBiMod} for filling fractions $f = 1/36$
(left) and $f=1/18$ (right). Such systems have a large density
gradient close to the edge, but is on the other hand more uniform in
the inner parts than the harmonic case. The effect is a scenario which
fits in between the cylindrical case and the harmonically trapped
system.  There are slight angular distortions of the outermost
vortices for $f=1/36$, and an increased possibility for vortex lattice
melting in the center as in an extended uniform system.
\par
\begin{figure}[htbp]
  %% \vspace{.5cm}
  \centerline{\hbox{\includegraphics[width=85mm]{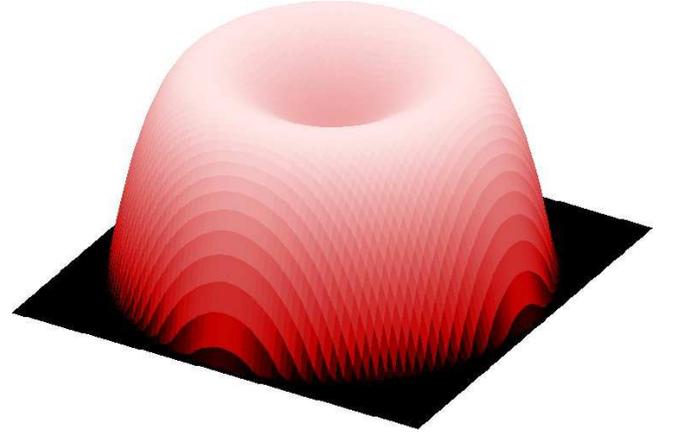}}}
  \caption{(Color online) Radial density profile $P_{i\mu}$ for the
    case of an anharmonic trap with a local minimum of the ground
    state condensate density at the center of the trap.}
  \label{fig:anharmonictrap}
\end{figure}
\begin{figure*}[htbp]
%%  \vspace{.5cm}
  \centerline{\hbox{\includegraphics[width=\textwidth]{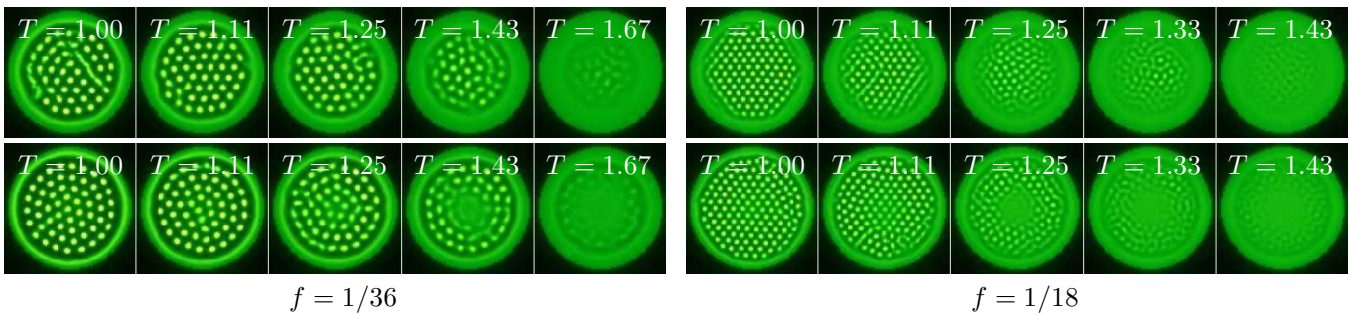}}}
  \caption{(Color online) $xy$-positions of the vortices in a pure
    quartic trapping potential (top row) with filling fractions $f =
    1/36$ (left) and $f = 1/18$ (right). In the second row, the
    vortices are trapped by a harmonic plus quartic potential, with
    $\alpha = 2$ in Eq. (\ref{eq:anharmdensity}). Due to lower
    density, the visibility of the vortices close to the rotation axis
    is reduced for the highest temperatures. All images are averages
    over 100 000 Monte Carlo sweeps.}
    \label{fig:vortexpositionsBiMod}
\end{figure*}

The generic ground state density profile $P_{i\mu}$ for the anharmonic
trap is shown in Fig. \ref{fig:anharmonictrap}, where the parameter
$\alpha$ in Eq. (\ref{eq:anharmdensity}) is taken to be $\alpha=2$,
such that the density profile has a distinct dip in the trap centre.

\par
In the second row of Fig. \ref{fig:vortexpositionsBiMod}, we have
chosen $\alpha=2$ so that the density profile has a local minimum in
the center. For low temperatures, the vortex configuration is close to
that of the harmonic system, see
Fig. \ref{fig:vortexpositionsharmonic}. Then we notice a decreased
visibility of the central vortices from around $T =1.25$, a feature
encountered in an experiment reported by Bretin and coworkers in 2004
\cite{bretin}, where they increased the rotation frequency above the
trap frequency. They speculated that this could be explained by
bending effects of the vortices. Our computations support this view.

\begin{figure}[htbp]
%%  \vspace{.5cm}
  \centerline{\hbox{\includegraphics[width=85mm]{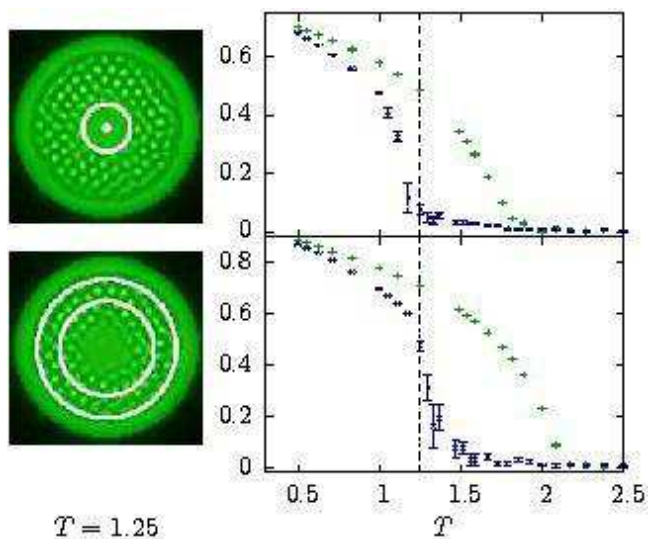}}}
  \caption{(Color online) Results for $\tilde{\Upsilon}_z(R_1,R_2)$,
    comparing the low- and high-density regions in the top and bottom
    panels respectively. The lower curve ($\times$) in both panels, as
    well as the images on the left, are calculated for a system with
    filling fraction $f=1/18$ and anharmonic density profile defined
    by $\alpha = 2$ (see Eq. (\ref{eq:anharmdensity})). Also included,
    are results for the same density profile, but with no
    rotation-induced vortices (upper curve, ($+$)).}
    \label{fig:FullHeliBiMod}
\end{figure}

Fig. \ref{fig:FullHeliBiMod} gives examples of the modified helicity
modulus $\tilde{\Upsilon}_z(R_1,R_2)$ for two distinct regions of a
system with $f = 1/18$ and $\alpha = 2$: One covering the region
closest to the rotation axis, and one around the peak in the
density. We have also included results for a non-rotating system, but
with the same density profile. The vortex lines clearly forces
$\tilde{\Upsilon}_z(R_1,R_2)$ to vanish in both regions for lower
temperatures than without rotation, indicating a melting of the vortex
line lattice. Additionally, we see that the region with the highest
density remains in the vortex solid state up to higher $T$ than the
inner part. The central vortex line lattice melts easier because in
this region there is a larger tendency for the vortex lines to have
bending fluctuations due to the higher effective temperature, in the
sense of Eq. (\ref{eq:effectiveT}).

\section{\label{sec:conclusion} CONCLUSIONS}

In this paper, we have considered thermal fluctuations of vortex
matter in cylindrical traps with varying non-uniform model ground
state densities of the condensate. As a benchmark we have also
performed Monte Carlo computations of a uniform system and a
cylindrical geometry with uniform ground state condensate density.  We
have in all cases taken care in ensuring that the helicity moduli in
the direction {\it transverse} to the direction of rotation have
vanished at temperatures well below the temperatures where the
helicity moduli {\it along} the direction of rotation vanish. The
purpose of this is to mimick, as well as possible, phase transitions
in the BEC-vortex system {\it in a continuum} and to eliminate the
artificial pinning effects that are inevitably present due to the
numerical grid.  Note however, that the model defined on a numerical
grid could have a physical realization in terms of rotating
Bose--Einstein condensates on an optical lattice. In such a system
there is a real frustration of the vortex system due to commensuration
effects. While this presents a very interesting problem in its own
right, a study of such effects has not been the purpose of this paper.
\par
{\it{\underline{Uniform system}}}: In a uniform system, the 3D $XY$
specific heat anomaly which the system exhibits in the absence of
rotation, is rounded to a broad non-critical peak when rotation is
present. This is the superfluid/BEC counterpart of the well-known
finite-field remnant of the zero-magnetic field superconductor--normal
metal transition when a supercondutor is subjected to a finite
magnetic field. In principle, it is also possible to extract a
specific heat anomaly associated with the melting of the BEC-vortex
lattice, although this has not been done in the present paper. The
vortex fluctuations of importance in this case are transverse phase
fluctuations of the superfluid order parameter.
\par
{\it{\underline{Uniform cylindrical confinement}}}: In a cylindrical
confinement with uniform ground state density profile, we have
investigated the thermal fluctuations of the ordered vortex lattice in
the confinement geometry as the temperature is increased. We find that
the vortex lattice is first excited close to the circular edge of the
confinement, and that the fluctuations in the vortex positions are
directed along the perimeter of the cylinder. As the temperature is
increased further, these ring-like thermal excitations creep inwards
towards the center of the cylinder until the vortex lattice also at
the center crosses over to a liquid phase. It is important that the
inhomogeneity observed in the vortex fluctuations in this case is not
due to the fact that the condensate effectively is warmer 
(in the sense of Eq. (\ref{eq:effectiveT}))
at the edge
than at the center. The effective temperature is uniform throughout the
system. 
The reason for the observed inhomogeneity of the fluctuations
is that the local environment around each vortex is different at the
edge than at the center. In particular, the vortex-vortex correlation
effects that impede thermally assisted vortex motion are smaller at
the edge than at the center. The coordination number of each vortex in
the vortex lattice is larger at the center than at the edge. The
corresponding interaction energies, and hence Coulomb barrier, that
must be overcome to produce vortex motion, is therefore smaller at the
edge of the confinement than at the center.  Moreover, a transverse
(angular) motion-pattern of the vortices are preferred compared to a
radial motion-pattern, for the following reason. In a superfluid/BEC,
the vortex interactions are unscreened (anti) Biot-Savart
interactions. Therefore, the vortex ensemble constitutes an {\it
  incompressible} system in a superfluid/BEC. Hence, vortex motion is
a collective process involving many vortices that have to be `pushed'
out of the way in order to pave the way for one vortex.  Motion along
the perimeter may be realized by moving all vortices in the {\it
  outermost ring} collectively, while radial motion involves a
rearrangement of {\it all} vortices in the system. The former is
obviously a collective process that is easier to accomplish than the
latter. As a corollary, we may infer that this ring-like excitation
pattern in a cylindrical confinement may change in the case of a {\it
  moderate type-II superconductor}, where effective screening of
vortex interactions is a consequence of a fluctuating gauge field.
This results in a compressible vortex system which to a much larger
extent will allow radial as well as angular fluctuations in vortex
positions.
\par
{\it{\underline{Harmonic and anharmonic traps}}}: For a Bose--Einstein
condensate in a magnetic trap, we have focused on two types of
trapping potentials, namely i) a harmonic trap giving a Thomas-Fermi
ground state density profile of the BEC, and ii) an anharmonic trap in
which we could vary the relative weights between a quartic and a
quadratic term. The latter trap naturally leads to a modification of
the inverse-parabolic Thomas-Fermi density profile, and in particular
it is possible to induce a ground state superfluid density with a
local minimum at the center of the trap. The thermally driven
vortex-excitations in these systems are fundamentally different from
the case of a uniform cylinder. The reason is that the vortex matter in
the trapped BECs
effectively have a highly non-uniform temperature, in the sense of
Eq. (\ref{eq:effectiveT}). This is due to the rapidly decreasing density
profile close to the edge, with a maximum effective temperature
gradient at the edge of the trap. Hence, this promotes vortex-loop
excitations as the edge of the trap is approached, thus inducing a
vortex-matter phase where the line tension (free energy per unit
length) of the vortex lines has vanished and the vortex lines
effectively have lost their directionality in the direction of the
rotation vector of the system. This gives a distinct region at the
outer edge of the trap where vortex-loop induced violent vortex-line
fluctuations wash out the image of each individual rotation-induced
vortex line.  This contrasts sharply with the images of ring-like
collective directed vortex-line excitations in the uniform
cylinder. The main difference between the harmonic and anharmonic trap
is that for a harmonic trap the effective temperature of the system
decreases monotonously towards the center of the trap, such that a
monotonous evolution of the system from a vortex tangle to a vortex
lattice at the center is observed when heated. For an anharmonic trap
with a dip in the ground state condensate density at the center of the
trap, we effectively have a local increase in the temperature of the
system as the center of the trap is approached. Hence, we can have a
tensionless vortex tangle at the center of the trap as well as at its
edge. Our computations therefore provide  a finite-temperature extension 
of the zero-temperature phase diagram for
anharmonic traps presented in Ref. \onlinecite{kavoulakis1}. According
to Refs. \onlinecite{kavoulakis2,kavoulakis3}, the experiment by
Bretin et al. \cite{bretin} takes place in the same regime, in which
the density and interaction strength are too high to obtain a giant
vortex with multiple quantization. We believe the experiment can be
described by our model.
\par
To summarize, we studied effects of density inhomogeneity and
finite-size in a model system describing a trapped Bose--Einstein
condensate. Although finiteness of the system and density
inhomogeneity prohibit use of the notion of a true phase transition
between different aggregate states of vortex matter (known to occur in
three dimensional extended model systems), we find that nonetheless
one can have various quasi-states of vortex matter surviving in a
rather robust form at finite length scales in traps.  Recent progress
in Bose--Einstein condensates has resulted in the availability of
systems with various symmetries and multiple components where one can
anticipate new states of vortex matter. This study suggests that
predictions of a theory based on a uniform density might nonetheless
have rather robust finite-size realizations in actual inhomogeneous
trapped systems.

\section*{Acknowledgments}
This work was supported by the Research Council of Norway, Grant Nos.
158518/431, 158547/431 (NANOMAT), 167498/V30 (STORFORSK), the National
Science Foundation, Grant No. DMR-0302347, Nordforsk Network on
Low-Dimensional Physics. SK and EB acknowledges the hospitality
of the Center for Advanced Study at the Norwegian Academy of Science
and Letters, where part of this work was done. 
%Discussions and correspondence with J.
%Dalibard, A. K. Nguyen, V. Schweikhard, E.  Sm\o rgrav, M. Zwierlein
%and especially with Erich J. Mueller, are gratefully acknowledged.


\begin{thebibliography}{99}


\bibitem{Coddington_Thesis} I. R. Coddington, {\it Vortices in a
    Highly Condensed Bose Gas}, PhD Thesis, University of Colorado,
  (2004).

\bibitem{Greiner_Thesis} M. Greiner, {\it Ultracold quantum gases in
    three-dimensional lattice potentials}, PhD thesis, Ludwig
  Maximilians Universit{\"a}t Muenchen, Germany (2003).

\bibitem{Regal_Thesis} C. Regal, {\it Experimental realizations of BCS-BEC
crossover physics with a Fermi gas of atoms}, PhD thesis, University 
of Colorado (2005). 
  
\bibitem{Zwierlein_Thesis} M. W. Zwierlein, {\it High-Temperature
    Superfluidity in an Ultracold Fermi Gas}, PhD Thesis,
  Massachusetts Institute of Technology, (2006).
  
  
\bibitem{KT} J. M. Kosterlitz and D. J. Thouless, J. Phys. C {\bf 6},
  1181 (1973); J. M. Kosterlitz, J. Phys. C {\bf 7}, 1084 (1974); {\it
    ibid}, {\bf 10}, 3753 (1977).
  
\bibitem{Hoye_Olaussen} For a thorough and comprehensive statistical
  mechanical treatment of the metal-insulator transition in the $2$D
  Coulomb gas, with particular emphasis on the difficult problem of
  treating the screening properties in the critical region, see
  J. S. H{\o}ye and K.  Olaussen, Physica {\bf 104A}, 447 (1980); {\it
    ibid}, {\bf 107A}, 241 (1981).
  
\bibitem{Kleinert1982} H. Kleinert, Lett. Nuovo Cimento, {\bf 35}, 409
  (1982).
  
\bibitem{Tesanovic1999} Z. Tesanovic, Phys. Rev. B {\bf 51}, 16204
  (1995); {\it ibid}, B {\bf 59}, 6449 (1999).
  
\bibitem{NguyenSudbo1998} A. K. Nguyen and A. Sudb{\o}, Phys. Rev. B
  {\bf 57}, 3123 (1998); {\it ibid}, B {\bf 58}, 2802 (1998).
  
\bibitem{NguyenSudbo1999} A. K. Nguyen and A. Sudb{\o}, Phys. Rev. B
  {\bf 60}, 15307 (1999); Europhys. Lett. {\bf 46}, 780 (1999).  See
  also A. K. Nguyen, R. E. Hetzel, and A. Sudb{\o}, Phys. Rev. Lett.,
  {\bf 77}, 1592 (1996).

\bibitem{Kleinert_book} H. Kleinert, {\it Gauge Fields in Condensed
    Matter Physics, Vol. 1}, World Scientific, Singapore 1989.

\bibitem{Fossheim_Sudbo_book} K. Fossheim and A. Sudb{\o}, {\it
    Superconductivity: Physics and Applications}, John Wiley \& Sons,
  Ltd (2004), see Chapters 9,10.


\bibitem{Onsager_vortexloop} The 1949 discussion remark of L. Onsager
  is well worth quoting here: "Finally, we can have vortex rings in
  the liquid and the thermal excitation of He II, apart from phonons,
  is presumably due to vortex rings of molecular size. As a possible
  interpretation of the $\lambda$-transition, we can understand that
  when the concentration of vortices reaches the point where they form
  a connected vortex tangle throughout the liquid, then the liquid
  becomes normal." L. Onsager, Nuovo Cimento Suppl., {\bf 249 },
  (1949). This idea was first put on a quantitative footing in $2$D in
  Refs. \onlinecite{KT}, and in $3$D in
  Refs. \onlinecite{Kleinert1982,Tesanovic1999,NguyenSudbo1998,NguyenSudbo1999}. In
  particular, see Fig. 3 of A. K. Nguyen and A. Sudb{\o}, Phys. Rev
  {\bf B} 60, 15304 (1999).

  
\bibitem{HoveSudbo} J. Hove, S. Mo, and A. Sudb{\o}, Phys. Rev. Lett., {\bf
    85}, 2368 (2000).

\bibitem{PWABasicNotions} P. W. Anderson, {\it Basic Notions in
    Condensed Matter Physics}, Benjamin Cummings, London (1984).

\bibitem{Ketterle_Nobelwork} W. Ketterle, K. B. Davis, M. A. Joffe,
  A. Martin, and D. E. Pritchard, Phys. Rev. Lett., {\bf 73}, 2253
  (1995); K. B. Davis, M. -O. Mewes, M. R. Andrews, N. J. van Druten,
  D. S. Durfee, D. M. Kurn, and W. Ketterle, Phys. Rev. Lett. {\bf
    75}, 3969 (1995)

\bibitem{Cornell_Nobelwork} W. Petrich, M. H. Anderson, J. R. Ensher,
  and E. A. Cornell, Phys. Rev. Lett., {\bf 74}, 3352 (1995);
  M. H. Anderson, J. R. Ensher, M. R. Matthews, C. E. Wieman, and
  E. A. Cornell, Science, {\bf 269}, 198 (1995).

\bibitem{Rice_work} C. C. Bradley, C. A. Sackett, and R. G. Hulet,
  Phys. Rev. Lett., {\bf 78}, 985 (1997).

\bibitem{BECVL1} M. R. Matthews, B. P. Anderson, P. C. Haljan,
  D. S. Hall, C. E. Wieman, and E. A. Cornell, Phys. Rev. Lett. {\bf
    83}, 2498 (1999).

\bibitem{BECVL2} C. Raman, M. K{\"o}hl, R. Onofrio, D. S. Durfee,
  C. E. Kuklewicz, Z. Hadzibabic, and W. Ketterle,
  Phys. Rev. Lett. {\bf 83}, 2502 (1999).

\bibitem{1dPRL} M. Snoek and H. T. C. Stoof, Phys. Rev. Lett. {\bf
    96}, 230402 (2006).

\bibitem{1dPRA} M. Snoek and H. T. C. Stoof, Phys. Rev. A {\bf
    74}, 033615 (2006).

\bibitem{Mueller} M. M. Parish, S. K. Baur, E. J. Mueller, and
  D. A. Huse, {\tt arXiv:0709.1120}, (2007).

\bibitem{NWA2004} E. Babaev, A. Sudb{\o}, and N. W. Ashcroft, Nature,
  {\bf 431}, 666 (2004).

\bibitem{Kragset} S. Kragset, E. Babaev, and A. Sudb{\o},
  Phys. Rev. Lett., {\bf 97}, 170403 (2006).

\bibitem{Leggett_review} A. J. Leggett, Rev. Mod. Phys., {\bf 73}, 307
  (2001), and references therein.

\bibitem{Brandt_Review} E. H. Brandt, Prog. Phys. {\bf 58}, 1465
  (1995).

\bibitem{Greiner1} M. Greiner, C. A. Regal, and D. S. Jin, Nature {\bf
    426}, 537 (2003).


\bibitem{Greiner2} C. A. Regal, M. Greiner, D. S. Jin,
  Phys. Rev. Lett. {\bf 92}, 040403, (2004).

\bibitem{Zwierlein1} M. W. Zwierlein, C. H. Schunck, A. Schirotzek,
  and W. Ketterle, Nature, {\bf 442}, 54 (2006).


\bibitem{Zwierlein2} Y. Shin, M. W. Zwierlein, C. H. Schunck,
  A. Schirotzek, and W. Ketterle, Phys. Rev. Lett. {\bf 97}, 030401
  (2006).


\bibitem{Zwierlein3} C. H. Schunck, M. W. Zwierlein, A. Schirotzek,
  and W. Ketterle, Phys. Rev. Lett. {\bf 98}, 050404 (2007).

\bibitem{Schunck2007} C.H. Schunck, Y. Shin, A. Schirotzek,
  M.W. Zwierlein, and W. Ketterle, Science, {\bf 316}, 867 (2007).

\bibitem{3DXY1} R. E. Hetzel, A. Sudb\o, and D. A. Huse, Phys. Rev.
  Lett. {\bf 69}, 518 (1992).

\bibitem{3DXY2} T. Chen and S. Teitel, Phys. Rev. B {\bf 55}, 15197
  (1997).

\bibitem{3DXY3} X. Hu, S. Miyashita, and M. Tachiki, Phys. Rev.
  Lett. {\bf 79}, 3498 (1997).

\bibitem{3DXY4} P.Olsson and S. Teitel, Phys. Rev.  Lett. {\bf 80},
  1964 (1998).

\bibitem{3DXY5} S. Ryu and D. Stroud, Phys. Rev. B {\bf 57}, 14476
  (1998).

\bibitem{TF1} G. Baym and C. J. Pethick, Phys. Rev. Lett., {\bf 76}, 6
  (1996).

\bibitem{TF2} G. Watanabe, G. Baym, and C. J. Pethick,
  Phys. Rev. Lett., {\bf 93}, 190401 (2004).

\bibitem{TF3} N. R.  Cooper, S. Komineas, and N. Read, Phys. Rev. A
  {\bf 70}, 033604 (2004).

\bibitem{Madison} K. W. Madison, F. Chevy, W. Wohlleben, and
  J. Dalibard, Phys. Rev. Lett., {\bf 84}, 806 (2000).

\bibitem{Coddington1} I. Coddington, P. Engels, V. Schweikhard, and
  E. A. Cornell, Phys. Rev. Lett., {\bf 91}, 100402 (2003).

\bibitem{Schweikhard} V. Schweikhard, I. Coddington, P. Engels,
  V. P. Mogendorff, and E. A. Cornell, Phys. Rev. Lett., {\bf 92},
  040404 (2004).

\bibitem{Smith} N. L. Smith, W. H. Heathcote, J. M. Krueger, and
  C. J. Foot, Phys. Rev. Lett., {\bf 93}, 080406 (2004).

\bibitem{Coddington2} I. Coddington, P. C. Haljan, P. Engels,
  V. Schweikhard, S. Tung, and E. A. Cornell, Phys. Rev. A {\bf 70},
  063607 (2004).

\bibitem{Muniz} S. R. Muniz, D. S. Naik, and C. Raman, Phys. Rev. A
  {\bf 73}, 041605 (2006).

\bibitem{metropolis} N. Metropolis, A. W. Rosenbluth,
  M. N. Rosenbluth, A. H. Teller, and E. Teller, J. Chem. Phys., {\bf
    21}, 1087 (1953).

\bibitem{hastings} W. K. Hastings, Biometrika, {\bf 57}, 97 (1970).

\bibitem{footnote_gridpinning} The resulting phase where one has zero
  helicity moduli in the direction transverse to the rotation vector,
  concomitant with a finite helicity modulus parallell to the rotaton
  vector, may be dubbed a ``floating solid phase'', since the vortex
  lattice ``floats'' as a solid while being thermally depinned from
  the numerical grid. Going to low enough filling fraction to avoid
  artificial pinning of the vortex lattice to the numerical grid, is a
  minimum requirement for studying phase transitions in vortex
  lattices in continuum systems. There are, however, interesting
  physical situations where one really has a rotating BEC defined on a
  discrete lattice system, most notably an optical lattices.  (See for
  instance, H. Pu, L. O. Baksmaty, and N. P. Bigelow,
  Phys. Rev. Lett., {\bf 94}, 190401 (2005); S. Tung, V. Schweikhard,
  and E. A. Cornell, Phys. Rev. Lett., {\bf 97}, 240402 (2006).)  In
  this case, one has competing length scales in the system, with
  associated competing energy scales. While this poses an extremely 
  interesting problem, both in $2$D as well as in $3$D systems, it is 
  beyond the scope of the present work.

\bibitem{Hasenbusch} M. Campostrini, M. Hasenbusch, A. Palisetto,
  P. Rossi, and E. Vicari, Phys. Rev B {\bf 63}, 214503 (2001).
   

\bibitem{Abrikosov} For a comprehensive review on the Abrikosov vortex
  lattice in type-II superconductors in a magnetic field, see
  G. Blatter, M. V. Feigel'man, V. B. Geshkenbein, A. I. Larkin, and
  V. M. Vinokur, Rev. Mod. Phys., {\bf 66}, 1125 (1994). 
%%  The Abrikosov vortex lattice appears as a consequence of flux
%%  quantization in type-II superconductors where the surface energy
%%  between a superconducting and normal region is negative.  In a
%%  superfluid or a BEC the same vortex structure appears as a result of
%%  quantization of rotation. In essence, the vortex lattice in
%%  superfluids has the same properties as the vortex lattice in extreme
%%  type-II superconductors such as high-$T_C$ superconductors. In
%%  type-II superconductors, there is an important distinction between
%%  flux lines and vortex lines. Flux lines are lines of confined
%%  magnetic flux with radius given by the magnetic penetration length
%%  $\lambda$. Vortex lines are lines of zeroes of the superconducting
%%  order parameter with radius given by the superconducting coherence
%%  length $\xi$. In moderate type-II superconductors, these lines
%%  essentially represent the same degrees of freedom, since $\lambda
%%  \approx \xi$. However, in extreme type-II superconductors, where
%%  $\lambda \gg \xi$, the typical length scale for bending of flux
%%  lines is vastly different from that of vortex lines. In the case of
%%  $\lambda/\xi \to \infty$ (the superfluid limit), flux lines are
%%  infinitely fat tubes of magnetic flux and are as such irrrelevant,
%%  while vortex lines still matter.

%%\bibitem{Volovik} G. Volovik, {\it The Universe in a Helium Droplet},
%%  Clarendon Press, Oxford (2003).

\bibitem{baym} S. A. Gifford and G. Baym, Phys. Rev. A {\bf 70},
  033602 (2004).

\bibitem{zipf} L. J. Campbell and R. M. Ziff, Phys. Rev. B {\bf 20},
  1886 (1979).

\bibitem{2dm} Yu. Lozovik and E. Rakoch, Phys. Rev. B {\bf 57}, 1214
  (1998).
  
\bibitem{bretin} V. Bretin, S. Stock, Y. Seurin, and J. Dalibard,
  Phys.  Rev. Lett. {\bf 92}, 050403 (2004).

\bibitem{Abo-Shaer} J. R. Abo-Shaeer, C. Raman, and W. Ketterle, Phys.
  Rev. Lett. {\bf 88}, 070409 (2002).
  
\bibitem{Ketterle} J. R. Abo-Shaeer, C. Raman, J. M. Vogels, and
  W. Ketterle, Science {\bf 292}, 476 (2001).

\bibitem{kavoulakis1} G. M. Kavoulakis and G. Baym, New Journal of
  Physics {\bf 5}, 51.1 (2003).

\bibitem{kavoulakis2} A. D. Jackson, G. M. Kavoulakis, and E. Lundh,
  Phys. Rev. A {\bf 69}, 0536619 (2004).

\bibitem{kavoulakis3} A. D. Jackson and G. M. Kavoulakis, Phys. Rev. A
  {\bf 70}, 023601 (2004).


\end{thebibliography}
\end{document}